\documentclass{osa-article}

\journal{osac}

\articletype{Research Article}

\begin{document}

\title{Optical vortex coronagraph imaging of a laser-induced plasma filament}

\author{Qingqing Liang,\authormark{1} Xia Huang,\authormark{1}  Yanfei Mou,\authormark{1} Shaodong Zhou,\authormark{1} Wenxing Zhang,\authormark{1} Jieyu Gui,\authormark{1} Grover A. Swartzlander, Jr.,\authormark{2} Qingqing Cheng,\authormark{1} and Yi Liu\authormark{1,3,*}}

\address{\authormark{1}Shanghai Key Lab of Modern Optical System, University of Shanghai for Science and Technology, 
516, Jungong Road, 200093 Shanghai, China
\\
\authormark{2}Chester F. Carlson Center for Imaging Science, Rochester Institute of Technology, Rochester, NY, USA\\
\authormark{3}CAS Center for Excellence in Ultra-intense Laser Science, Shanghai, 201800, China}

\email{\authormark{*}yi.liu@usst.edu.cn} 



\begin{abstract}
A high contrast imaging technique based on an optical vortex coronagraph (OVC) is used to measure the spatial phase profile induced by an air plasma generated by a femtosecond laser pulse.  The sensitivity of the OVC method significantly surpassed both in-line holographic and direct imaging methods based on air plasma fluorescence.  The estimated phase sensitivity of 0.046 waves provides opportunities for OVC applications in areas such as bioimaging, material characterization, as well as plasma diagnostics.
\end{abstract}

\section{Introduction}

Phase contrast imaging is an essential technique in modern biology for observations of transparent objects such as living cells, micro-sectioned tissue, and subcellular particles, with cross-over applications in many areas of material science \cite{dong2020super,lewis2004medical,faridian2013high} . In the classic Zernike scheme of phase contrast microscopy, an annual phase mask for phase manipulation of the un-scattering light is used in the Fourier plane, which leads to light intensity modulation proportional to the phase of the object formed in the imaging plane \cite{zernike1942phase}. With wide applications of the Zernike technique, several different methods of spatial manipulation in the Fourier plane have been proposed and implemented, such as Schlieren method by blockage of half of all the spatial frequency, and dark-field method by filtering of the low spatial frequency component \cite{faridian2013high,elias2018improving,marquet2005digital,pfeiffer2006phase}. More recently the filtering of light in the Fourier plane by a helical phase (vortex) mask attracted considerable interest in the imaging community\cite{khonina1992phase,hsu1982rotation,jaroszewicz1993zone,sacks1998holographic,davis2000image,mawet2005annular,fo2005optical,lee2006experimental,furhapter2007spiral,swartzlander2008astronomical,2009The,mawet2009optical,serabyn2010image,ritsch2017orbital}.  A vortex phase mask is characterized by a topological charge index, $m$, which performs an $m^{th}$ order Hankel transform on the image \cite{jaroszewicz1993zone,sacks1998holographic}. Fourier filtering using a vortex with $m=1$ provides a type of edge enhancement of both amplitude and phase objects and has found applications in image process \cite{hsu1982rotation,davis2000image,furhapter2007spiral,ritsch2017orbital,jesacher2006spiral} .Phase modulation in the Fourier plane with a vortex mask of $m = 2$ (and other non-zero even values) was found to provide extreme high contrast astronomical imaging for direct exoplanet imaging \cite{mawet2005annular,fo2005optical,lee2006experimental,swartzlander2008astronomical,2009The,mawet2009optical,serabyn2010image}  The optical configuration of the latter device, called an optical vortex coronagraph (OVC), is similar to a traditional Lyot coronagraph, but with the small opaque occulting disk replaced with a large transparent vortex phase mask.In view of the fact that a large “zero-valued interior” is formed in the imaging plane, some of the current authors have proposed that this optical coronagraph can also be used for background free imaging of a phase object and the detection sensitivity can be significantly enhanced \cite{Grover2006The}. Up to now, this possibility has not been exploited experimentally and its performance remains unknown. 

In this paper, we explore the use of an OVC for high contrast imaging of the weak phase disturbance produced by a laser-induced plasma filament. After verifying the proposed method by use of numerical simulations we report our experimental findings below. We compare the OVC method with the classic in-line holography method \cite{papazoglou2008line} and direct imaging of plasma fluorescence \cite{liu2005experiment}. It was found that the OVC technique reached a detection limit well beyond the other two methods. Moreover, we found that the length of the air plasma calibrated with the coronagraph and the holograph methods are significantly longer than that obtained with fluorescence imaging, which indicates that the commonly used fluorescence imaging method underestimates the plasma length. Numerical simulations of nonlinear pulse propagation were also conducted to estimate both the plasma density and the corresponding phase disturbance. 

\section{Optical Vortex Coronagraph Imaging Technique}

A schematic diagram of the OVC imaging system is depicted in Fig. 1. 
A circularly polarized collimated laser beam having a uniform irradiance profile is transmitted through a phase object or “sample”, 
which is located close to lens L1 of focal length $f_1 = 300 \, \mathrm{mm}$. 
An aperture stop AS with radius $R$ is placed adjacent to the lens, serving as an entrance pupil.  
A second lens, L2 of focal length $f_2 = 300 \, \mathrm{mm}$ is placed a distance $f_2$ from the vortex mask,
thereby forming an image of AS a distance $f_2 (f_1+f_2) / f_1$ from L2 (see Fig. 1(b)).
A second aperture called a Lyot stop (LS) of radius $R' \le f_2 R / f_1$ is placed in this image plane.
A vector vortex phase retarder with transmission function $\exp(i m \theta)$ and
topological charge $m = 2$  (VR2, Shenzhen Lubang Technology Co., Ltd.) 
is placed in the back focal plane of L1.  The transmission function of VR2 is given by $\exp(i m \theta)$.
When the mask is centered on the optical axis the image at the Lyot stop changes to a 
``ring of fire” as depicted in Fig. 1(c). 
The Lyot stop blocks the ring of light, transmitting only the dark inner region which is subsequently
imaged with unity magnification on a 12-bit sensor array (Newport model LBP2-HR-VIS2) by use of 
lens L3 having focal length $f_3 = 75 \, \mathrm{mm}$.  The sensor (CCD) has a pixel pitch of $3.69 \, \mathrm{\mu m}$.
A weak phase object in the sample region appears as a signal on an otherwise dark background.
This ``background-free" property of the OVC imaging system is an advantage not afforded by other
phase contrast imaging techniques such as in-line holographic and direct imaging methods based on air plasma fluorescence.

\begin{figure}[ht!]
\centering\includegraphics{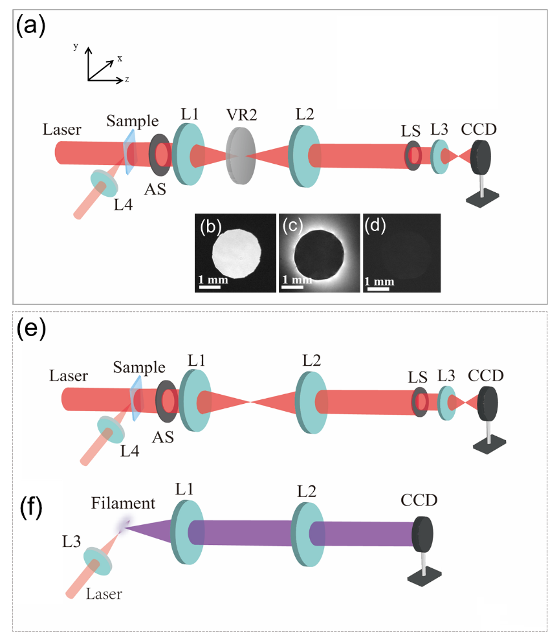}
\caption{(a) Optical vortex coronagraph schematic.  
A uniform planar probe beam is transmitted through the sample region.  
An aperture stop (AS) of radius $R$, placed adjacent to lens L1 of focal length $f_1$, 
is imaged on a Lyot stop (LS) of radius $R' = f_2 R / f_1$ with lens L2 of focal length $f_2$.  
The lenses are separated by $f_1 + f_2$ and LS is a distance $f_2 (f_1+f_2)/f_1$ from L2.
A lens L3 images LS on a CCD detector array where a uniform disk of light is observed (b).
Inserting a vortex phase mask in the back focal plane of L1 produces a characteristic
ring of fire (c) at LS and a black field on the sensor (d).
Lens L4 focuses a femtosecond laser in the sample, forming a plasma filament which
appears in the dark field.}
\end{figure}


\subsection{Background-free property of the OVC technique}

We assume the sample imposes a small phase $| \Phi (r,\varphi) | < 1$ on a probe beam
described by a uniform plane wave.  The beam is truncated at AS (see Fig. 1) 
by a circular pupil function $P(r) =1$ for $r < R$ and zero otherwise such that the field at the aperture
stop may be expressed

\begin{equation}{\label{E_AS}}
E_{\text {AS }} (r, \varphi) = P(r) e^{i \Phi (r,\varphi) } \approx P(r) + i P(r) \Phi(r,\varphi) 
\end{equation}

\noindent The electric field in the plane of the Lyot stop may be described using Fourier optics principles
and the convolution theorem \cite{2009The}:

\begin{equation}\begin{split}
{\label{e1}}
E_{\text {LS}}\left(r^{\prime}, \varphi^{\prime}\right) & = 
\mathrm{FT}^{-1}\left\{e^{i 2 \theta} \mathrm{FT} \left\{ E_{\text {AS}}(r, \varphi) \right\}\right\}
\\
& = P(r) \otimes \mathrm{FT}^{-1}\left\{ e^{i 2 \theta} \right\}  +
      i P(r) \, \Phi(r,\varphi) \otimes \mathrm{FT}^{-1}\left\{ e^{i 2 \theta} \right\}  
\end{split} \end{equation}

\noindent where $\mathrm{FT}$ represents Fourier transform,
$\otimes$ represents convolution, and $\mathrm{FT}^{-1} \left\{ e^{i 2 \theta} \right\} = -\exp(i 2 \varphi) / \pi r^2$.

Without a phase object in the sample region, e.g., $\Phi = 0$
a ``ring of fire" of radius $R' = f_2 R / f_1$ is formed at the Lyot stop:

\begin{equation}{\label{e2}}
E_{\text {LS }}\left(r^{\prime}, \varphi^{\prime}\right)=-e^{i 2 \varphi^{\prime}}\left\{\begin{array}{cc}
	0 \, ,  & r^{\prime} < R' \\
	\left(R' / r^{\prime}\right)^{2} \, , & r^{\prime} > R'
\end{array}\right.
\end{equation}

\noindent Therefore the inner region in the plane of the Lyot stop, $r'  < R'$, only contains 
light scattered by the phase object:

\begin{equation}
{\label{E_LS_2}}
E_{\text {LS}}\left(r^{\prime}, \varphi^{\prime}\right) = 
      -\frac{i}{\pi} \Big( P(r) \, \Phi(r,\varphi) \Big) \otimes \left( \frac{e^{i 2 \varphi}} {r^2} \right) \, , \ \ r' < R'
\end{equation}

\subsection{Simulation of optical vortex coronagraph imaging}

It is clear that the light intensity in the imaging plane is related to the phase of the object, providing the base of this technique for imaging of a phase object. We numerically simulated this imaging process and the corresponding results are presented in Fig. 2. The form of the object “panda” is 
\begin{figure}[ht!]
	\centering\includegraphics{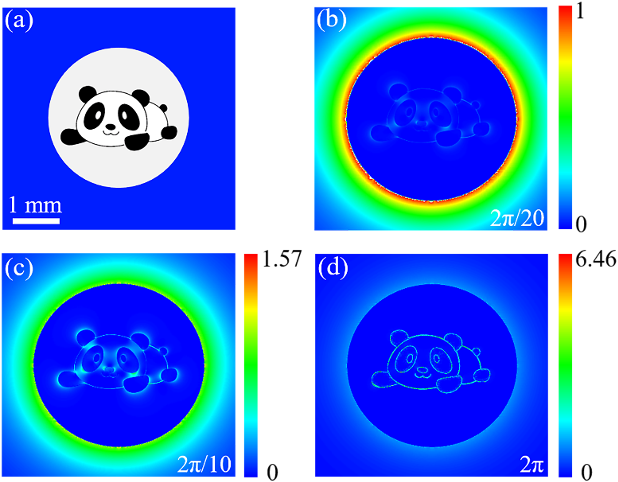}
	\caption{(a) Input phase object for the numerical simulation of the coronagraph imaging. (b)-(d) Simulated OVC images of the phase object in (a), with different phase changes denoted in the panels. }
\end{figure}
shown in Fig. 2(a) and the simulation imaging results for different phase variation is presented in Fig. 2 (b)-(d). The images well reproduce the form of the phase object for phase variations ranging from $2 \pi / 20$ to $2 \pi$.

\section{Experimental results and discussion}

\subsection{Experiment setup}

To demonstrate the experimental validation of the optical coronagraph technique, we employed an air plasma induced by femtosecond laser pulses as the phase object. Femtosecond laser pulses with pulse duration of 35 fs was focused by a convex lens of $f$ = 100 mm into ambient air. The incident beam diameter is 11 mm, which corresponds to a numerical aperture NA = 0.1. The input pulse energy was varied from tens of microjoule up to 1.6 mJ, by a variable neutral optical density. When the laser intensity in the focal area reaches the ionization threshold of $\sim 10^{13} W/cm^2$, the oxygen and nitrogen molecules become to be ionized and a dilute plasma is formed \cite{couairon2007femtosecond}. With the gradual increase of the pump pulse energy, the plasma density increases and tend to be saturated on the order of $10^{18}-10^{19} cm^{-3}$ in this case of NA = 0.1\cite{theberge2006plasma}. In the meantime, the length of the plasma increases significantly while its width presents a less sensitive dependence on the incident pulse energy \cite{mitryukovskiy2014backward}. It is well known that the plasma exhibits a negative refraction index of $\Delta \mathrm{n} = -\rho_{e} / 2 \rho_{c}$, where $\rho_{e}$ is the plasma density and 
$\rho_{c}=\varepsilon_{0} m_{e} \omega^{2} / e^{2}$ is the critical plasma density for a given frequency $\omega$ of the light \cite{couairon2007femtosecond}. Then, the phase accumulated by the imaging femtosecond beam propagating in the $z $direction after transmission the  air plasma is the integration of the refraction index change over the propagation distance, ie, $\frac{2 \pi}{\lambda} \int_{-\infty}^{+\infty} \Delta n(x, y, z) d z$.

To evaluate the sensitivity and detection limit of this optical coronagraph method, we compared it with the in-line holograph method, which is also commonly used for characterizing of the air plasma \cite{papazoglou2008line}. The in-line holograph technique can be readily implemented in our experiments by removing the phase mask of m = 2 in the Fourier plane, as presented in Fig. 1(e). Another method widely used to characterize the air plasma is imaging of its fluorescence by a camera in the visible light range, which has been widely used due to its extreme simplicity \cite{couairon2007femtosecond,mitryukovskiy2014backward}. The fluorescence of the air plasma stems mainly from the emission of the excited neutral and ionic nitrogen molecules, as well as the Bremstrlung radiation of free electrons inside the plasma \cite{talebpour1999re}. In our experiments, we also took imaging of the air plasma fluorescence with the same CCD used for the coronagraph imaging. In this case, the imaging femtosecond laser beam is blocked and the vortex mask is also removed, as shown in Fig. 1(f).

\subsection{Experimental results and discussion}

\begin{figure}[ht!]
	\centering\includegraphics{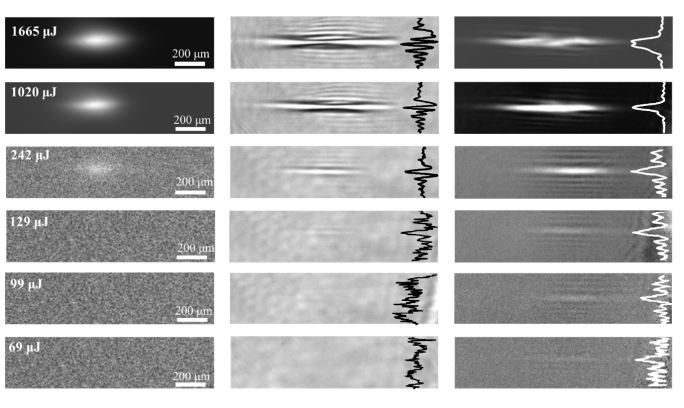}
	\caption{Left column: direct fluorescence images of the air plasma pumped by femtosecond laser pulses. The energies of the incident pump laser pulses are presented in the corresponding panels. Middle column: the diffraction patterns. Right column: the images obtained by the OVC technique.}
\end{figure}

In Fig. 3 (left column), the images of the air plasma in the visible light range recorded by the CCD are presented for different incident pulse energies. For the highest pulse energy of 1.6 mJ, the plasma image presents a length of 0.32 mm (Full-Width at Half Maximum, FWHM) and a width of 0. 11 mm. With the current setup, a barely visible image of the plasma was
obtained for pulse energy of 242 $\mu $J, below which the image is buried in the noise of the CCD device. In the middle column of Fig. 3, the results for in-line holograph method are presented. The incident pulse energies are the same as that of the left column. Holograph of the plasma with clear intensity modulation are observed for incident pulse energy above 129 $\mu $J, which should be considered as the pulse energy threshold for this method. From these holograph
patterns, the plasma density can be retrieved by properly assuming the geometrical configuration of the plasma object \cite{papazoglou2008line}. The fine structures of the diffraction pattern will be discussed in details later.

\begin{figure}[ht!]
	\centering\includegraphics{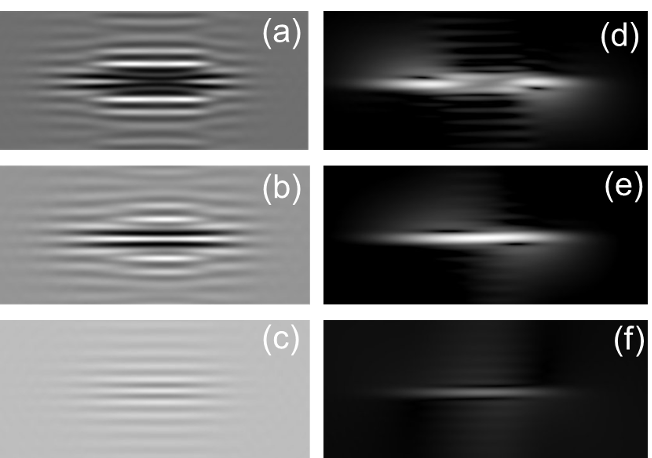}
	\caption{Simulated holograph patterns and OVC images for different phase accumulation $\varphi _0$ by the imaging beam. (a)-(c) holograph patterns, (d)-(f) images of the OVC technique.}
\end{figure}

The optical coronagraph images are presented in the right column of Fig. 3 for the same pump energy energies. It is noticed that an image can be obtained for incident pump pulse energy as low as 70 $\mu $J, going significantly beyond the direct imaging and the holograph methods, which demonstrates that the optical vortex coronagraph method has a much better detection sensitivity compared to the other methods. For higher incident pulse energy, the coronagraph image consists of a bright central stripe surrounded by fringes. We noticed that for incident pump energy of 1.67 mJ the central bright stripe split around its center and a local minimum develops. This is obvious in the crossline cut presented as the inset of the top-right panel of Fig. 3.

To get further insight into the coronagraph images (right column, Fig. 3) and holograph patterns (middle column, Fig. 3), we performed numerical simulations of these two methods. In the simulation, we assume that the air plasma exhibits density distribution of Gaussian form in both its longitudinal and transverse directions. The plasma density reads as:  $\rho(x, r)=\rho_{0} e^{-\left(\frac{x}{\omega_{l}}\right)^{2}-\left(\frac{r}{\omega_{r}}\right)^{2}}$. Here $\omega_{r}$ and $\omega_{l}$are the width of the distribution in the transverse and longitudinal directions. As to the parameters of the simulations, the length of the air plasma can be determined from the images in the right column of Fig. 3, while the width of the plasma was assumed to be 30 $\mu $m for NA = 0.1, according to the previous reports in the literature \cite{theberge2006plasma}. The other parameters concerning the geometrical configuration of the setup are taken from our experiments. The only free parameter in the simulation is the maximum plasma density $\rho_0 $.

In Fig. 4, we presented the simulated holograph patterns and the coronagraph images for typical plasma densities. These results should be compared with the middle and right columns of Fig. 3. Concerning the formation of the holograph patterns, it is observed that some symmetrical thinner fringes appears around the central zone when the accumulated phase of the imaging beam exceeds $\pi$ (Fig. 4 (a)). This is in good agreement with the experimentally observed diffraction patterns. Next, we turn to the images of the optical vortex coronagraph. For accumulated phase far less than $\pi$ , the images consist of bright central stripe with surrounding horizontal fringes, in good agreement with the experimental observations. With the increase of the phase $\varphi _0$ ,the central bright stripe presents splitting inclined 45$^\circ$ with respect to the horizontal direction. The coronagraph image on top-right of Fig. 3 is well reproduced when the phase reaches over $\pi$. 

\begin{figure}[ht!]
	\centering\includegraphics{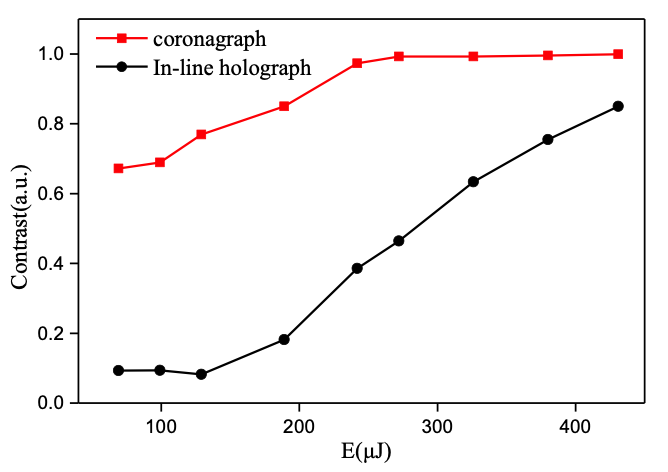}
	\caption{Contrast of the OVT imaging and the holograph patterns as a function of the pump laser pulse energy.}
\end{figure}

To quantitatively compare the optical vortex coronagraph method with the holograph method, we defined the contrast parameter as: $\gamma=\left(I_{\max }-I_{\min }\right) /\left(I_{\max }+I_{\min }\right)$. The contrast of both methods extracted from experimental results are presented in Fig. 5. For incident energy around the detection limit of the diffraction method, the contrast of optical coronagraph method is about 10 times better than that of the diffraction patterns. For increased pump pulse energy, the diffraction pattern and the coronagraph image both becomes more visible and the contrast increases. The contrast for the coronagraph method can reach nearly 100\% for incident pump pulse energy above 0.26 mJ, thanks to its background free property. 

\subsection{Numerical simulation for the air plasma formation and further discussion}

To evaluate the phase detection limit of the optical coronagraph technique, we performed numerical simulation of the nonlinear propagation of the 70 $\mu$J femtosecond laser pulse in ambient air with the code SWIFT \cite{couairon2011practitioner}. The parameters for simulation are taken from the experiments, including those of the pump pulse and the focusing condition. From the numerical simulations, we can obtain the evolution of the plasma density and the laser intensity along the propagation direction x of the pump laser. In this case of relatively tight focusing geometry (NA = 0.1) with pulse peak power far below the critical power for self-focusing ($P_{cr}$ = 3.4GW), it was found that the pulse propagates almost linearly and the laser intensity reaches $7\times10^{15}W/cm^{2}$ at the focus. With such high laser intensity, the air molecules are fully ionized and the maximum plasma density is thus equal to the density of air molecules in ambient air, i.e, $\rho _0 = 2.5\times10^{19}/cm^3$. The phase experienced by the probe beam is then estimate to be $\varphi(x, y)=\frac{2 \pi}{\lambda_{0}} \int_{-w_{0} / 2}^{+w_{0} / 2} \Delta n(x=0, y=0, z) d z=2 \pi / 21.7$, considering the transverse width of the plasma at the focus is calculated to be around  $\sim 5\mu$m (FWHM).

\begin{figure}[ht!]
	\centering\includegraphics{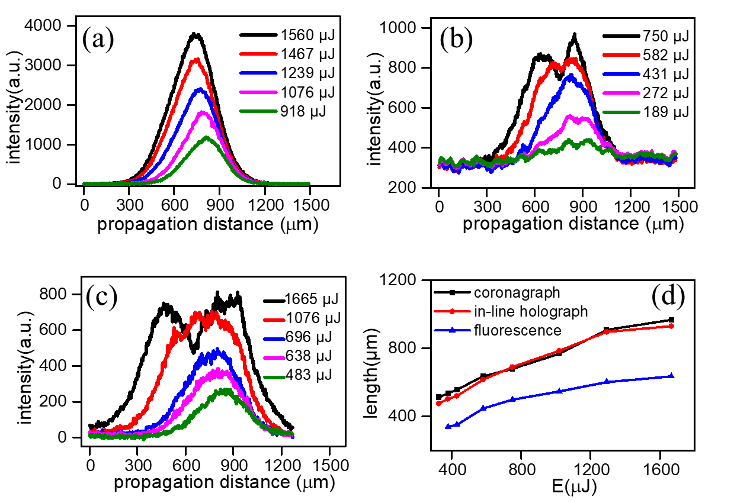}
	\caption{Distribution of the fluorescence intensity (a), the intensity of diffraction pattern (b), and the intensity of the OVC images (c) along the x direction (the propagation direction of the pump laser). (d), Comparison of the air plasma length determined by the three different methods.}
\end{figure}

Finally, we would like to discuss one difference that we noted in Fig. 3. One notices that the plasma length captured by the direct imaging method (left column) is obviously shorter than that of the diffraction method and the coronagraph technique. In Fig. 6 (a)-(c), the horizontal cross-section line of all the three methods are presented for different incident laser pulse energies. The length of the air plasma determined by the three methods are presented in Fig. 6 (d). The plasma length determined by the coronagraph and holograph methods agree well with each other, while those determined by the direct imaging method of plasma fluorescence are systematically smaller. We attributed this difference to the fact that both the coronagraph technique and the holograph method are sensitive to the phase accumulation of the imaging beam through the plasma, which is linear proportional to the density of the electron $\rho_{e}$. In contrast, the direct imaging of air plasma is based on the fluorescence emission of the excited neutral molecules in the $C^{3} \Pi_{u}^{+}$ state and ionic nitrogen molecules in the $B^{2} \Sigma_{u}^{+}$ state \cite{talebpour1999re}. The excited neutral nitrogen molecules $\mathrm{N}_{2}\left(C^{3} \Pi_{u}^{+}\right)$ are mainly formed via collision excitation of the nitrogen molecules in the ground state by energetic electron with kinetic energy above the threshold energy of 14 eV \cite{danylo2019formation,mitryukovskiy2015plasma}. For the excited ionic nitrogen molecules $N_{2}^{+}\left(B^{2} \Sigma_{u}^{+}\right)$, they origin from the ionization of the inner electron in the HOMO-2 orbit of $N_2$ \cite{talebpour1999re}, which obviously requires much higher laser intensity with respect to that of HOMO orbit. As a result, it can be expected that the plasma length determined by the fluorescence method is shorter, since it relies on excited $N_2$ and $N_{2}^{+}$which require relatively higher laser intensity. 

\section{Conclusion }

In summary, we demonstrate that the optical vortex coronagraph technique based on the phase manipulation in the Fourier plane with a vortex mask of m = 2 can serve as a novel method for imaging of a phase object. The unique advantage of this technique is that a background-free image is formed, which is confirmed both theoretically and experimentally. We employ an air plasma induced by femtosecond laser pulses as a phase object and compare the optical vortex coronagraph technique with the commonly used holograph method and the direct imaging of the air plasma fluorescence. It was found that the OVC technique exhibits a much better detection limit regarding pump pulse energy as low as 70 $\mu$J, going much beyond the other two methods. The phase detection limit was estimated to be $\lambda / 21.7$ by numerical simulation of the nonlinear interaction of the femtosecond laser pulse with the ambient air. If femtosecond laser pulse like herein are used as the imaging beam, this method can be easily implemented in pump-probe experiments for time-resolved measurement. We believe that this simple and sensitive imaging technique for phase object may find applications ranging from bioimaging to ultrafast material characterization dynamics, plasma diagnostic, $etc$.

\begin{backmatter}
\bmsection{Funding}
The work is supported by the National Natural Science Foundation of China (Grants No. 11904232, 11874266, 12034013), Innovation Program of Shanghai Municipal Education Commission (Grant No. 2017-01-07-00-07-E00007). 

\bmsection{Disclosures}

\end{backmatter}

\bibliography{sample}






\end{document}